\pdfoutput=1
\documentclass[manuscript]{rmaa}
\usepackage {graphicx}
\usepackage{natbib}
\def\aap{A\&A\,  }
\def\aaps{A\&AS  }
\def\apj{ApJ\,  }
\def\apjl{ApJ\,  }
\def\apjs{ApJS  }
\def\apss{Astrophysics and Space Science  }

\def\mnras{MNRAS\,  }
\def\nat{Nature\,  }

\def\physa{Phys. A    }
\def\za{Z. Astrophys.  } 
\title
{
Chord distribution along a line in the local Universe
}
\author{L. Zaninetti\altaffilmark{1}}

\altaffiltext{1}{Dipartimento di Fisica, Via Pietro Giuria 1,
10125 Torino,  Italy (zaninetti@ph.unito.it.)}
\shortauthor{Zaninetti}
\shorttitle {Statistics of the voids}
\ReceivedDate{...............}
\AcceptedDate{...}
\SetYear{2012}
\resumen{Este documento describe ..
 }

\abstract
{
A method is developed  to compute  the
chord length distribution along  a line
which intersects  a cellular  Universe.
The cellular  Universe is here modeled
by the Poissonian Voronoi Tessellation (PVT)
and by  a non-Poissonian Voronoi Tessellation (NPVT).
The distribution of the spheres is obtained
from common approximations used in modeling
the  volumes of Voronoi Diagrams.

We give analytical formulas  for the distributions of the lengths of chords in both the PVT and NPVT.
The astrophysical  applications are made to the real
Eso Slice Project  and to an artificial slice
of galaxies which simulates
the 2dF Galaxy Redshift Survey.
}
\addkeyword{statistical distributions }
\addkeyword{galaxies }
\addkeyword{clusters of galaxies}

\begin{document}
\maketitle

\section{Introduction}

\begin{figure*}
\begin{center}
\includegraphics[width=7cm]{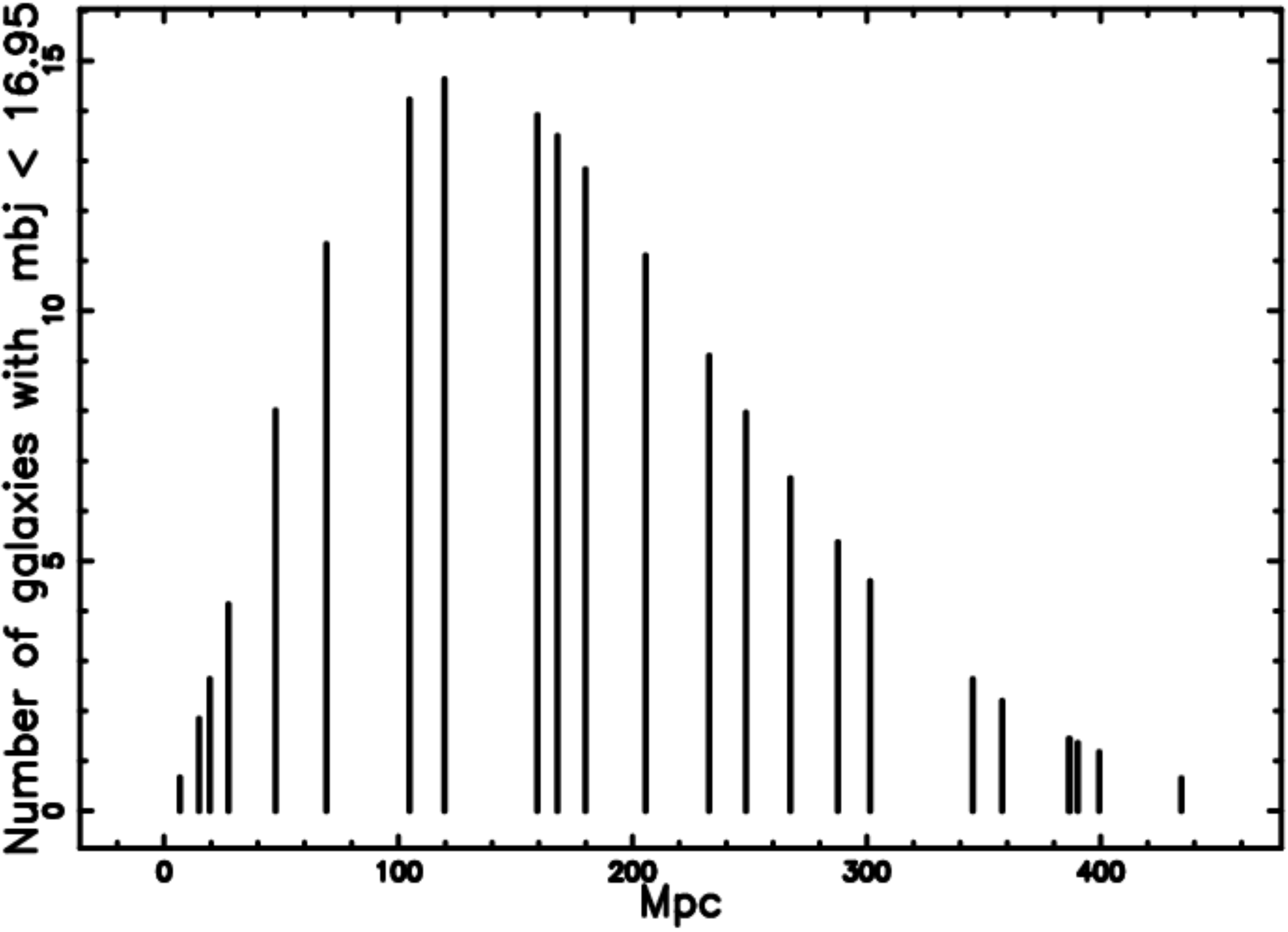}
\end {center}
\caption{
The number of galaxies  as  a function  
of the distance  for  a pencil beam 
catalog  of $\Omega = 60$  deg$^2$.
The curve  was calibrated on the data of   the 
2dF Galaxy Redshift Survey (2dFGRS)
which has  $\Omega=1500$ deg$^2$.
Adapted from Figure 4  in  \citet{zaninetti2010a}.
}
          \label{introduction}%
    \end{figure*}

The study of the spatial periodicity
in the spatial  distribution of quasars  started
with  \citet{Tifft1973,Tifft1980,Tifft1995}.
A recent analysis  of  \citet{Bell2006},
in which  46 400  quasars were processed,
quotes   a periodicity  near  $\Delta z$ =0.7.
The  study  of periodicity  in
the spatial distribution of galaxies
started with \citet{broadhurst},
where  the data  from four distinct surveys at the north and
south Galactic poles were processed.
He found an apparent
regularity in the galaxy distribution
with a characteristic scale
of 128  Mpc.
Recently, \citet{Hartnett2009a,Hartnett2009b}  quotes
peaks  in the distribution of galaxies
with a periodicity  near  $\Delta z$ =0.01.
More precisely, he found a
regular real space radial
distance spacings of 31.7 Mpc, 73.4   Mpc, and 127 Mpc.
On adopting a theoretical  point of view,
the periodicity is  not easy  to  explain.
A framework which explains the periodicity
is given by the cellular  universe
in which the galaxies are situated  on the faces
of irregular  polyhedrons.
A reasonable model for the cellular universe
is the Poissonian Voronoi
Tessellation (PVT);
another is the non-Poissonian Tessellation (NPVT).
Some  properties  of the  PVT can be  deduced by introducing
the averaged radius  of a polyhedron, $\bar{R}$.
The  astronomical  counterpart is  the  averaged
radius of  the cosmic voids  as given  by  
the Sloan Digital Sky Survey (SDSS)
R7, which  is $\bar{R}= \frac{18.23}{h} Mpc$, 
see \citet{Vogeley2012}.  
The  number of intercepted  voids, $n_v$, 
along  a line  will  be  $n_v= \frac{L}{\bar{l}} $,
where  $L$ is  the considered length  and  $\bar{l}$ the average
chord. 
On assuming  $\bar{l} =\frac{4}{3} \bar{R} $,
$n_V = \frac{3L }{4 \bar{R}}$.
The  astronomical counterpart   of the  line is the pencil 
beam catalog, a cone characterized by a   narrow
solid angle, see \citet{szalay1993}.
The number of galaxies intercepted on the faces 
of the PVT  will  follow  the photometric rules
as  presented in  \citet{zaninetti2010a},
and  Figure \ref{introduction}  reports  the  theoretical  
number
of  galaxies for  a pencil  beam     
characterized by the solid angle $\Omega$ of 
60  deg$^2$.
According  to the cellular structure  of the local universe,
the number of galaxies  as a function of distance
will follow a discontinuous  rather than a continuous
increase or decrease.

We can therefore  raise the following questions.
\begin{itemize}
\item
Can  we find  an analytical  expression
for the  chord length distribution for lines
which intersect  many  PVT or NPVT polyhedrons?
\item
Can  we  compare  the observed periodicity
with  the theoretical  ones?
\end{itemize}
In order  to answer these questions,
Section~\ref{basic} briefly reviews
the existing knowledge of
the chord's length in the presence  of a given distribution
of spheres.
Section  \ref{voronoichords} derives  two new equations
for the chord's distribution in a PVT or NPVT environment.
Section  \ref{astrophysicalsec} is devoted to
a test of the  new formulas  against a real astronomical slice
and against a simulated slice.

\section{The basic equations}

\label{basic}

This Section reviews
a first  example
of deducing the average chord  of spheres
having the same diameter and
the general formula for
the chord's length  when a given distribution
for the spheres' diameters is given.

\subsection{The simplest example}

\begin{figure*}
\begin{center}
\includegraphics[width=10cm]{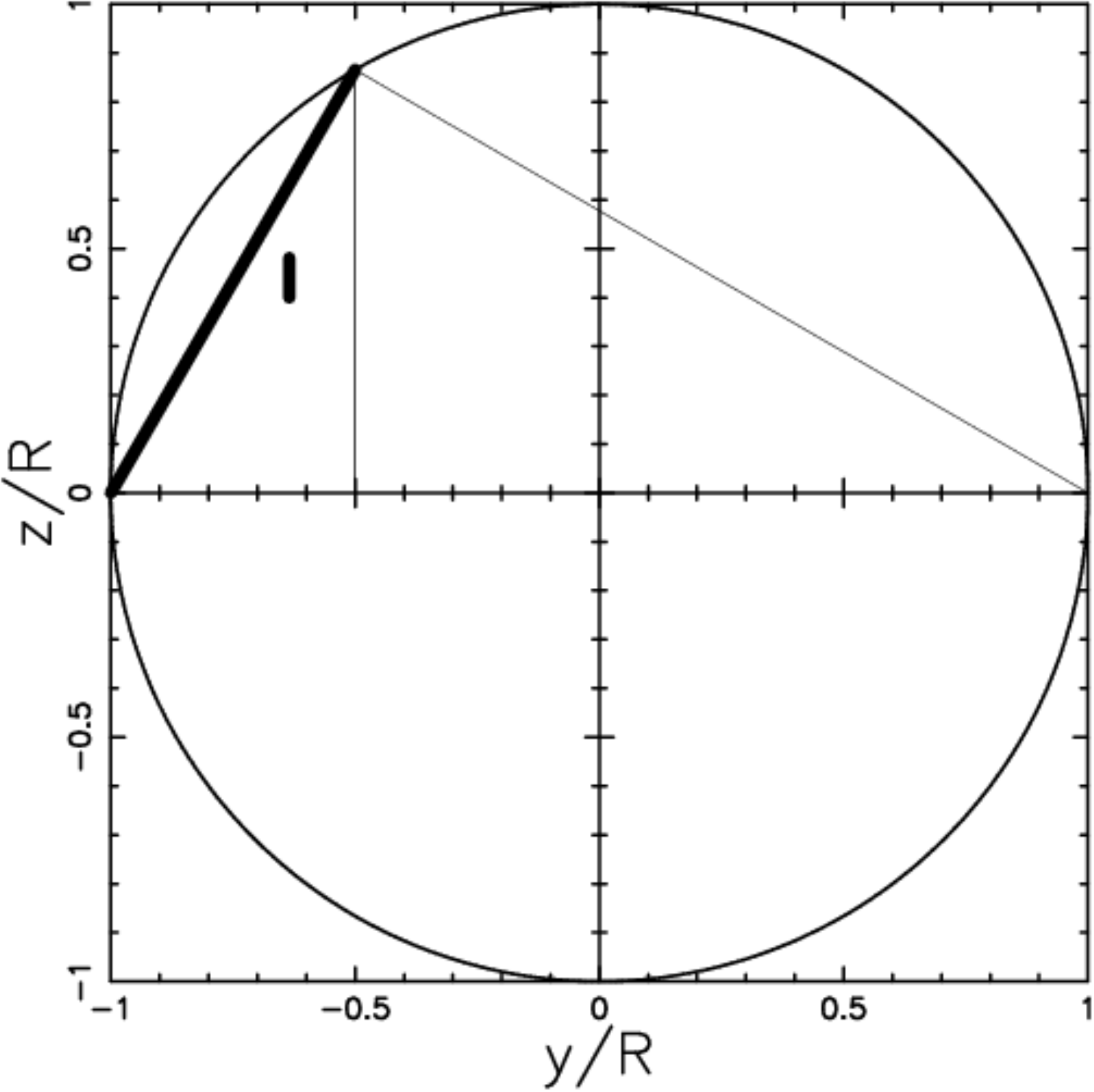}
\end {center}
\caption
{
The
section of an intersected sphere of unit radius.
The chord is  drawn   with  the thicker line  and
marked with $l$.
}
\label{chord_simple}
    \end{figure*}

We calculate  the average length  of all chords
of spheres  having the same radius $R$.
Figure \ref{chord_simple}
reports  a section of a sphere
having radius  $R$  and chord  length $l$.
The  Pythagorean theorem gives
\begin{equation}
l= \sqrt{2(1+y)} R
\quad  ,
\end{equation}
and the average  chord  length  is
\begin{equation}
<l> =
\frac{1}{2} \int_{-1}^{+1} \sqrt{2(1+y)}  R dy   = \frac{4}{3} R
\label{monogeometrical}
\quad  .
\end{equation}

\subsection{The probabilistic approach}

\begin{figure*}
\begin{center}
\includegraphics[width=10cm]{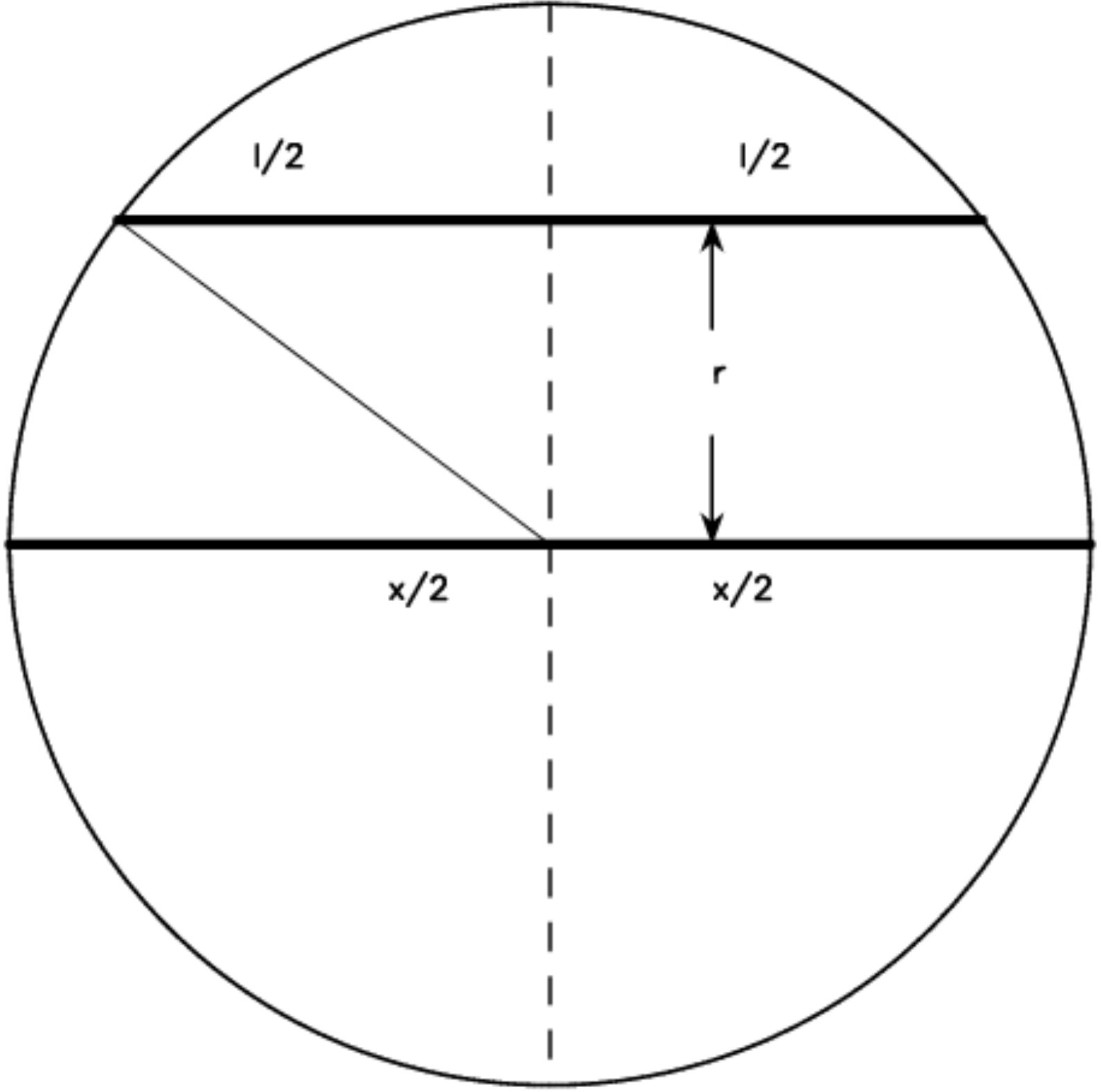}
\end {center}
\caption
{
The
section having diameter $x$
of the intersected sphere.
The chord is  drawn   with  the thicker line  and
marked with    $l$;
the distance  between  chord
and center is  $r$.
}
\label{chord_statistics}
    \end{figure*}

The  starting point is  a probability density function
(PDF)  for the diameter of the voids,
$F(x)$,  where   $x$  indicates  the diameter.
The  probability, $G(x)dx$,
that  a sphere having diameter between
$x$ and  $x+dx$ intersects  a random line is
proportional  to their  cross section
\begin{equation}
G(x) dx  =  \frac { \frac{\pi}{4} x^2  F(x) dx }
                  { \int_0 ^{\infty}
                    \frac{\pi}{4} x^2  F(x) dx}
=
\frac {  x^2  F(x) dx }
      {  < x^2>}
\quad  .
\end{equation}
Given a line  which intersects a sphere of diameter
$x$, the probability that the distance
from the center  lies  in the range
$r,r+dr$  is
\begin{equation}
p(r) = \frac{2 \pi r dr }{\frac{\pi}{4} x^2 }
\quad ,
\end{equation}
and the chord length is
\begin{equation}
l = \sqrt { x^2 - 4r^2}
\quad ,
\end{equation}
see  Figure \ref{chord_statistics}.

The probability that  spheres  in the
range  $(x,x+dx)$   are intersected to produce
chords  with lengths  in the range
$(l,l+dl)$ is
\begin{equation}
G(x)\, dx  \frac{2l\,dl}{x^2}
=
\frac{2l \, dl} { <x^2>}  F(x) dx
\quad .
\end{equation}
The probability  of having a chord
with  length  between $(l,l+dl)$  is
\begin{equation}
g(l)
=
\frac{2l} { <x^2>}  \int_l^{\infty} F(x) dx
\quad .
\label{fundamental}
\end{equation}
This integral  will be called {\it fundamental}
and the previous  demonstration  has  been  adapted
from \citet{Ruan1988}.
A first test of the previous  integral  can be done
inserting  as a distribution  for the diameters
a Dirac delta function
\begin{equation}
F(x)=\delta (x-2\,R)
\quad .
\end{equation}
As a consequence,  the following  PDF for chords
is obtained:
\begin{equation}
g(l)  = \frac{1}{2} \frac{l}{R^2}
\quad ,
\end{equation}
which has an average value
\begin{equation}
<l> =
\frac {4}{3} R
\quad  .
\end{equation}
We have therefore  obtained, in the framework
of the probabilistic approach, the same result
deduced with  elementary methods,
see Equation (\ref{monogeometrical}).

\section{Voronoi diagrams}

\label{voronoichords}
This section reviews
the  distribution of spheres  which approximates
the volume distribution  for
PVT and NPVT, explains  how to generate NPVT seeds, and   derives
two  new  formulas     for the distributions of the chords.

\subsection{PVT  volume distribution}

We analyze the gamma variate $H (x ;c )$  (\citet{kiang})
\begin{equation}
 H (x ;c ) = \frac {c} {\Gamma (c)} (cx )^{c-1} \exp(-cx),
\label{kiang}
\end{equation}
where $ 0 \leq x < \infty $, $ c~>0$,
and $\Gamma$ is the gamma function.
The Kiang  PDF has a mean of
\begin{equation}
\mu = 1,
\end{equation}
and variance
\begin{equation}
\sigma^2 = \frac{1}{c}.
\end{equation}
A new PDF due to  \citet{Ferenc_2007}
models the normalized area/volume
in  2D/3D PVT
\begin{equation}
FN(x;d) = C \times x^{\frac {3d-1}{2} } \exp{(-(3d+1)x/2)},
\label{rumeni}
\end{equation}
where $C$ is a constant,
\begin{equation}
C =
\frac
{
\sqrt {2}\sqrt {3\,d+1}
}
{
2\,{2}^{3/2\,d} \left( 3\,d+1 \right) ^{-3/2\,d}\Gamma \left( 3/2\,d+
1/2 \right)
},
\end{equation}
and $d(d=1,2,3)$ is the
dimension of the space under consideration.
We will call this
function the  Ferenc--Neda  PDF;
it has a mean of
\begin{equation}
\mu = 1,
\end{equation}
and variance
\begin{equation}
\sigma^2 = \frac{2}{3d+1}.
\end{equation}
The Ferenc--Neda  PDF  can be obtained from the Kiang function
(\citet{kiang}) by  the transformation
\begin{equation}
c =\frac{3d+1}{2},
\label{kiangrumeni}
\end{equation}
and  as an example  $d=3$ means  $c=5$.

\subsection{NPVT  volume distribution}

The most used   seeds
which produce  the tessellation
are the so  called
Poissonian seeds.
In this, the most explored  case, the volumes are
modeled  in 3D  by a Kiang function,
Equation  (\ref{kiang}), with  $c=5$.
An increase of the value of $c$ of the Kiang
function   produces  more ordered  structures
and a decrease, less ordered structures.
A careful  analysis  of the  distribution
in  effective radius of the Sloan Digital Sky Survey (SDSS)   DR7
indicates  $c \approx 2$, see  \citet{zaninetti2012e}.
Therefore the  normalized  distribution  in volumes
which  models the  voids between galaxies is
\begin{equation}
 H (x ;2  ) =4\,x{{\rm e}^{-2\,x}}
\quad  .
\label{kiangc2}
\end{equation}

\subsection{NPVT  seeds}
\label {secnewseeds}
The  3D set of seeds which generate a distribution in volumes 
with $c \approx 2$ for the Kiang function (\ref{kiangc2}) is produced
with the following algorithm.
A given number, $N_{H}$, of forbidden spheres  having   radius 
$R_H$ are generated in a 3D box  having side $L$.
Random seeds are produced  on the three spatial coordinates; those
that fall inside the forbidden spheres are rejected.
The volume forbidden to the seeds   occupies  the following 
fraction, $f$, of the total available  volume
\begin{equation}
f = \frac{N_H \frac{4}{3} \pi R_H^3 } {L^3}
\quad .
\end{equation}
The  value of   $c \approx 2$ for the Kiang function is found
by increasing progressively  $N_{H}$ and  $R_H$.

\subsection{PVT chords}

\begin{figure*}
\begin{center}
\includegraphics[width=10cm]{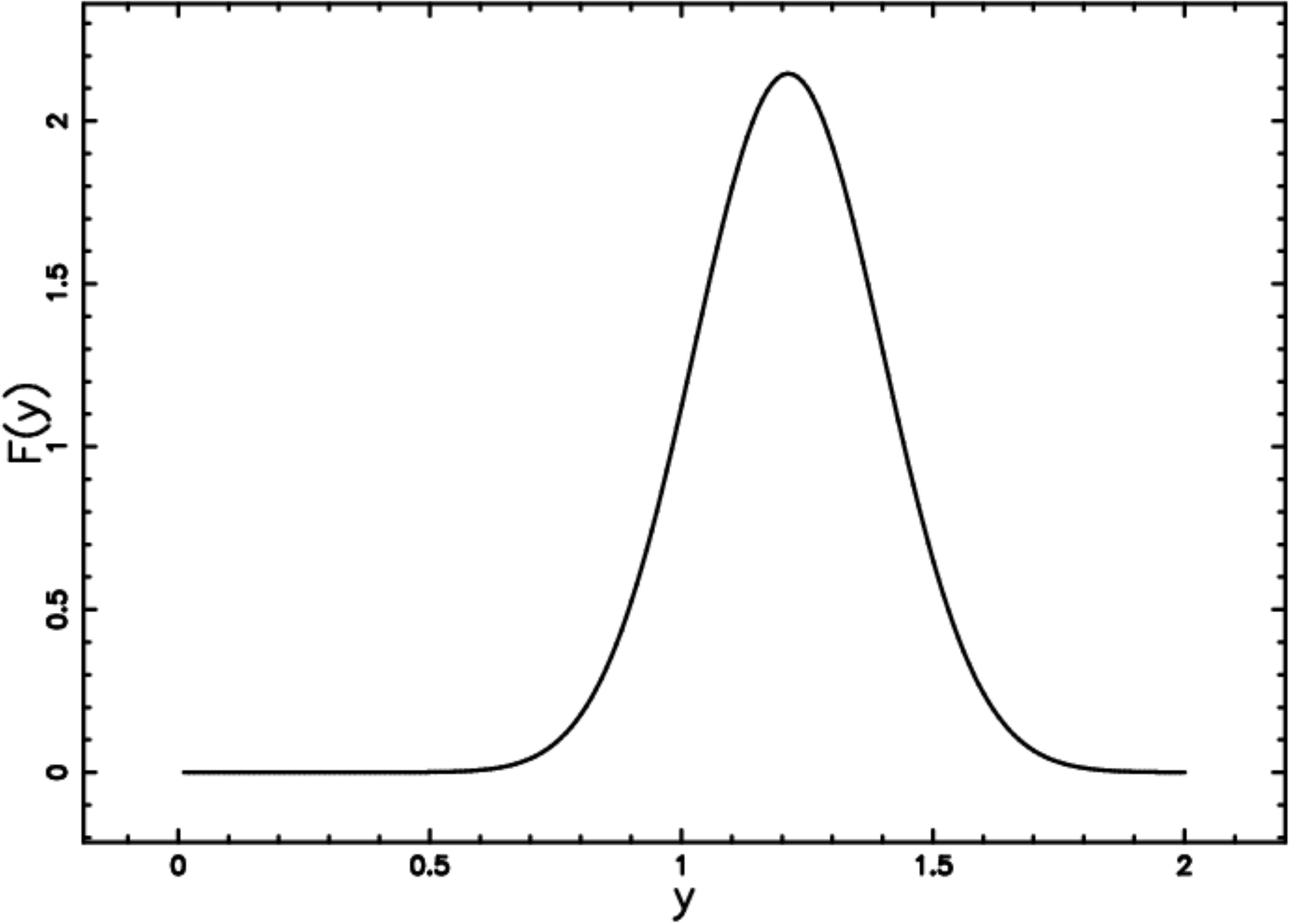}
\end {center}
\caption{
PDF  $F(y)  $ for the  PVT diameters.
}
\label{fxdiametri}
    \end{figure*}

The distribution in volumes  of the 3D PVT
can be modeled by the modern
Ferenc--Neda  PDF (\ref{rumeni}) by inserting $d=3$.
The corresponding  distribution in diameters
can be found
using the following  substitution
\begin{equation}
x  = \frac{4}{3} \pi (\frac{y}{2})^3
\quad  ,
\end{equation}
where  $y$ represents
the diameter of the volumes
modeled by the spheres.
Therefore the  PVT distribution in diameters  is
\begin{equation}
F(y)= {\frac {3125}{62208}}\,{\pi }^{5}{{\it y}}^{14}{{\rm e}^{-5/6\,\pi \,
{{\it y}}^{3}}}
\quad  .
\end{equation}
Figure ~\ref{fxdiametri} displays   the PDF of the  diameters
already obtained.

\begin{figure*}
\begin{center}
\includegraphics[width=10cm]{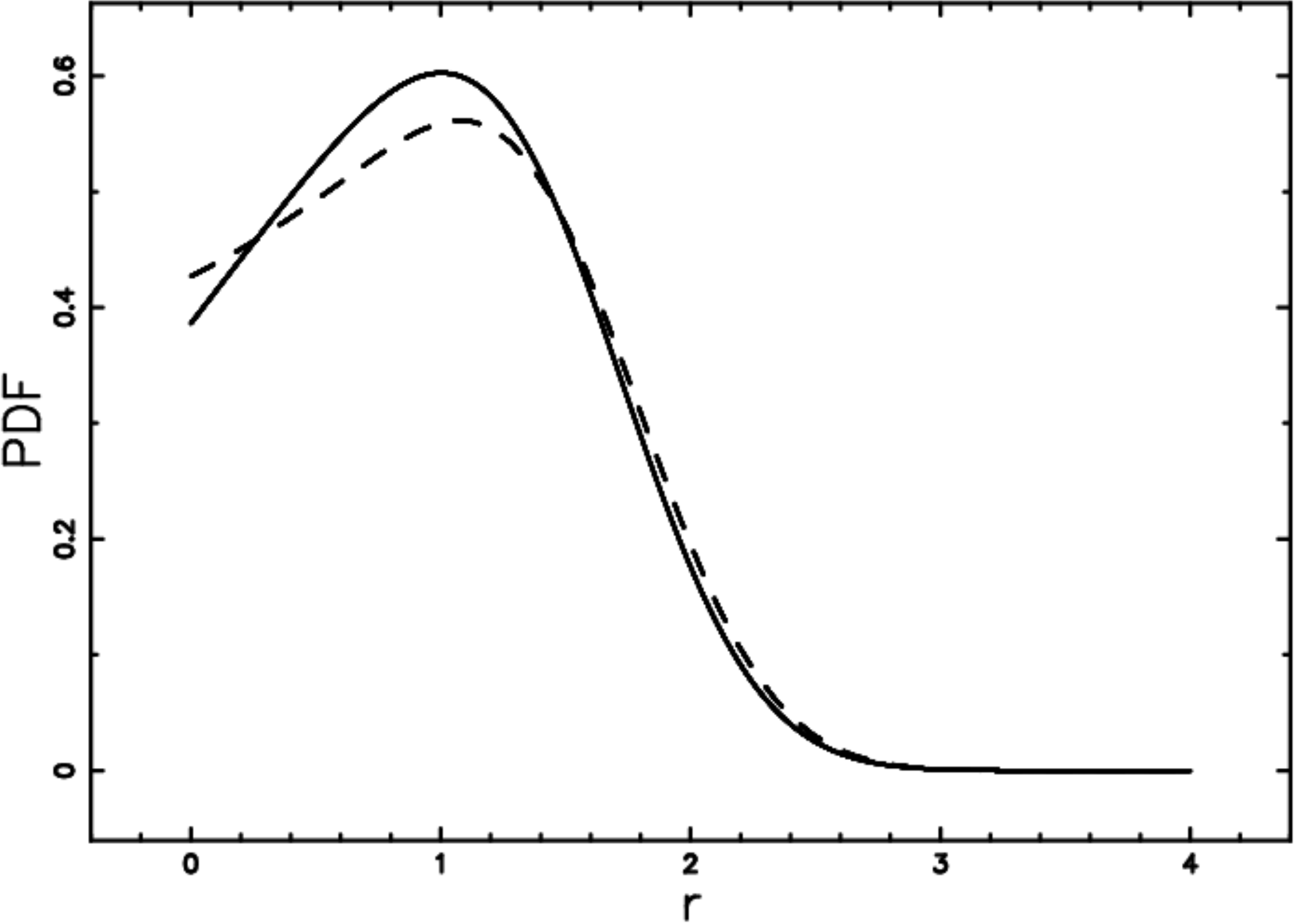}
\end {center}
\caption
{
PDF  $g_3 (r;a,b)  $ (full line) 
for chord length as a function of $r$
when $b=2.452$, $a=0.563$, which
means $<r> =1$,   and  the  mathematical PDF (dashed line)  
as extracted from Table 5.7.4  in Okabe et al. (2000),
PVT case.
}
\label{corda_pdf}
    \end{figure*}

We are now ready  to insert in the fundamental
Equation (\ref{fundamental}) for the chord  length
a PDF for the diameters.
The resulting integral  is
\begin{eqnarray}
g(l)=
{\frac {243}{1540}}\,{\frac {l{5}^{2/3}\sqrt [3]{6}{\pi }^{2/3}}{
\Gamma  \left( 2/3 \right)  \left( {{\rm e}^{\pi \,{l}^{3}}} \right) ^
{5/6}}}+{\frac {81}{616}}\,{\frac {{l}^{4}{5}^{2/3}\sqrt [3]{6}{\pi }^
{5/3}}{\Gamma  \left( 2/3 \right)  \left( {{\rm e}^{\pi \,{l}^{3}}}
 \right) ^{5/6}}}
\nonumber    \\
+{\frac {135}{2464}}\,{\frac {{l}^{7}{5}^{2/3}\sqrt [
3]{6}{\pi }^{8/3}}{\Gamma  \left( 2/3 \right)  \left( {{\rm e}^{\pi \,
{l}^{3}}} \right) ^{5/6}}}+{\frac {75}{4928}}\,{\frac {{l}^{10}{5}^{2/
3}\sqrt [3]{6}{\pi }^{11/3}}{\Gamma  \left( 2/3 \right)  \left( {
{\rm e}^{\pi \,{l}^{3}}} \right) ^{5/6}}}
\nonumber \\
+{\frac {125}{39424}}\,{
\frac {{l}^{13}{5}^{2/3}\sqrt [3]{6}{\pi }^{14/3}}{\Gamma  \left( 2/3
 \right)  \left( {{\rm e}^{\pi \,{l}^{3}}} \right) ^{5/6}}}
\end{eqnarray}
This first result should be corrected due to the fact that  the PDF
of the diameters, see Figure \ref{fxdiametri}, is  $\approx$ 0 up
to a diameter of  $\approx$ 0.563. We therefore introduce the
following translation
\begin{equation}
U = L+ a \quad ,
\end{equation}
where  $L$ is the  random chord  and  $a$ the amount of the
translation. The new translated PDF, $g_{1}(u;a)$, takes
values $u$ in the interval $[-a, (6-a)]$. Due to the fact that
only positive chords are defined in the interval $[0, (6-a)]$, a
new constant of normalization should be introduced
\begin {eqnarray}
g_{2}(u;a)= Cg_1(u;a) , \quad where \\
C=\frac{1}{\int_0^{6-a}g_1(u;a) du}
\quad .
\end{eqnarray}
The last transformation  of scale is
\begin{equation}
R = \frac{U}{b} \quad ,
\end{equation}
and the definitive  PDF for chords is
\begin{equation}
g_3(r;a,b)=
\frac{g_2(\frac{u}{b};a)}{b} \quad .
\label{GLBPOISSONIAN}
\end{equation}
The resulting distribution function will be
\begin{equation}
DF_{1,3}(r:a,b) =\int_0^r g_3(r;a,b) dr \quad .
\label{dfpoisson}
\end{equation}
We are now ready for a comparison with the 
distribution function
,$F_{L_{1,3}} $, for chord length 
, $L_{1,3}$ in $V_p(1,3)$, see
formula  (5.7.6) and Table 5.7.4  in \citet{Okabe2000}.
Table~\ref{table_parameters} shows the average diameter,
variance,  skewness,  and kurtosis of the already derived
$ g_3 (r;a,b) $.
The parameter $b$  
should  match the average value  of the  PDF in   
Table 5.7.4  of  \citet{Okabe2000}.
\begin{table}
 \caption[]
{
The parameters   of  \lowercase{$g_3 (r;a,b) $},
Eq.~(\ref{GLBPOISSONIAN}), relative to
the  PVT  case when 
\lowercase{$b=1.624$} and \lowercase{$a=0.563$}. 
}
 \label{table_parameters}
 \[
 \begin{array}{ll}
 \hline
Parameter       & value   \\ \noalign{\smallskip}
 \hline
 \noalign{\smallskip}
Mean            & 0.662   \\
\noalign{\smallskip}
\hline
Variance        & 0.144  \\
\noalign{\smallskip}
\hline
Skewness        &  0.274  \\
 \hline
Kurtosis        &  -0.581  \\
 \hline
 \end{array}
 \]
 \end {table}
The behavior  of $g_3(r;a,b)$ 
is shown in Figure  \ref{corda_pdf}.

\begin{figure*}
\begin{center}
\includegraphics[width=10cm]{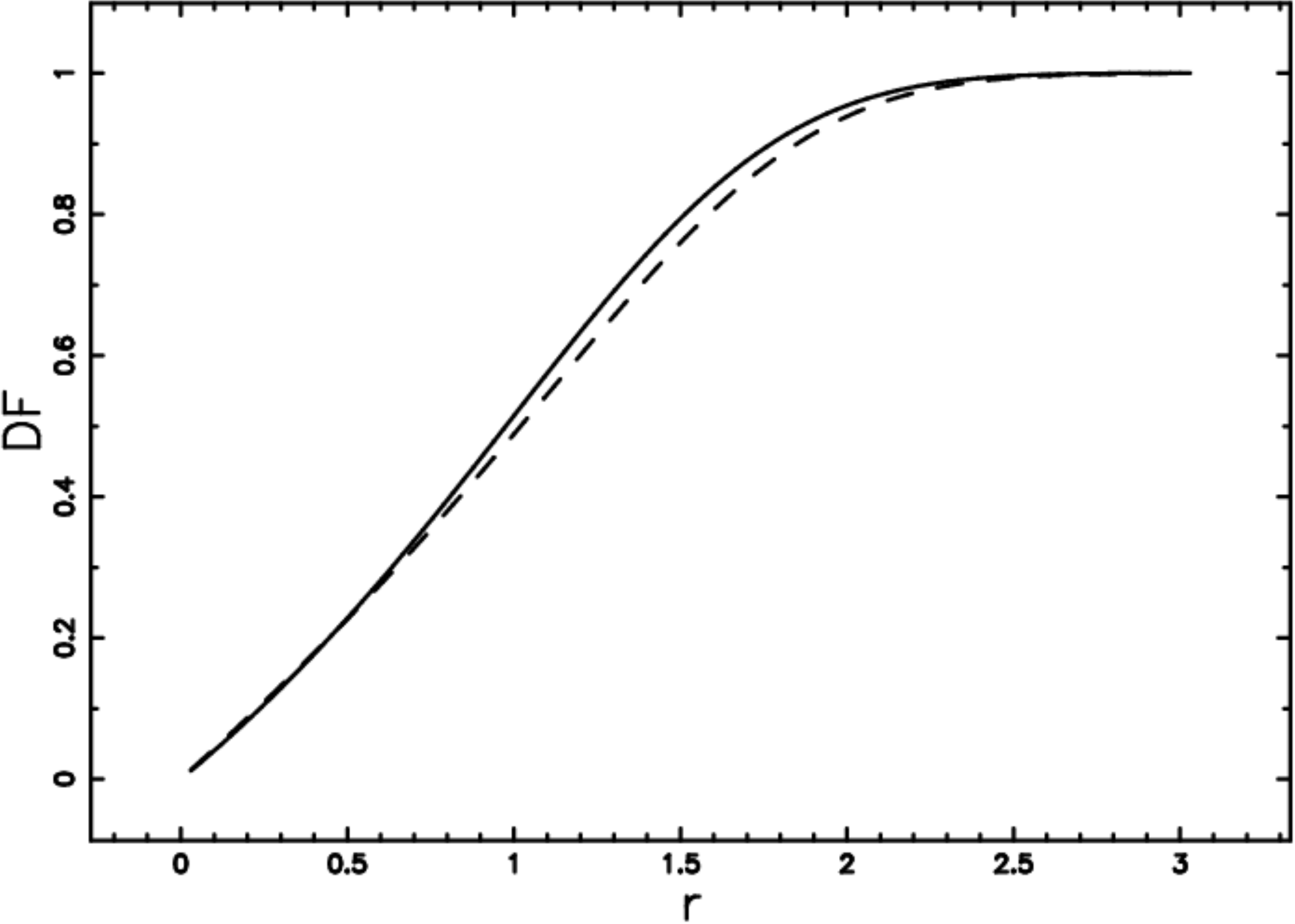}
\end {center}
\caption{
Distribution function  
$ DF_{1,3}(r:a,b) $ (full line) 
for chord length
as a function of $r$ when
$b=2.452$, $a=0.563$, which 
means $<r> =1$,
and  the  mathematical DF (dashed line)  
as extracted from Table 5.7.4  in Okabe et al. (2000),
PVT case.
}
\label{corda_df}
    \end{figure*}

\begin{figure*}
\begin{center}
\includegraphics[width=10cm]{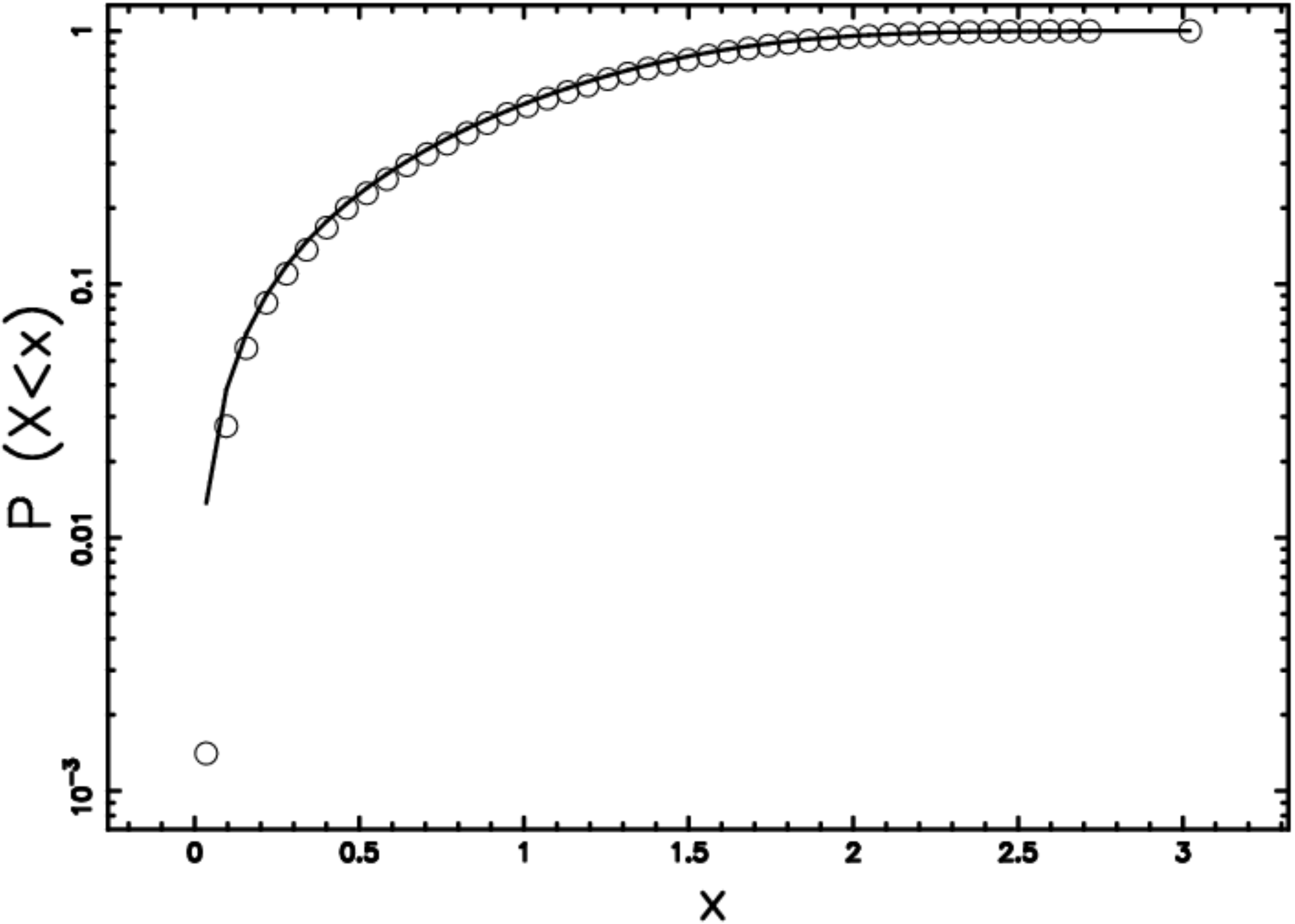}
\end {center}
\caption
{
Comparison between data
(empty circles) and theoretical curve , $ DF_{1,3}(r:a,b) $, 
(continuous line)
of  the chord length
distribution when
$b=2.452$, $a=0.563$, which 
means $<r>$ =1;  PVT  case.
The maximum distance  between the two curves
is $d_{max}=0.01$.
}
\label{corda_poisson}
    \end{figure*}

The behavior  of $DF_{1,3}(r:a,b) $
is shown in Figure  \ref{corda_df}.
Consider a  3D PVT  and suppose
it intersects a randomly oriented line $\gamma$:
the theoretical distribution function
$DF_{1,3}(r:a,b)$    as given by Eq. (\ref{dfpoisson})
and the results
of a numerical simulation  can be compared.
We start  from $900 000$ 3D
PVT  cells   and we process
$9947$ chords  which
were obtained by adding together  the results of
$40$ triples of  mutually perpendicular  lines.
The  numerical distribution of Voronoi lines  is
shown in Figure  \ref{corda_poisson}  with
the  display of  $DF_{1,3}(r:a,b)$.

\subsection{NPVT chords}

In the  case here analyzed, see the  PDF in volumes as
given  by  Equation(\ref{kiangc2}), the  NPVT distribution
in diameters  which corresponds to $c=2$ in volumes  is
\begin{equation}
F(y)=
1/3\,{\pi }^{2}{{\it y}}^{5}{{\rm e}^{-1/3\,\pi \,{{\it y}}^{3}}}
\quad  ,
\end{equation}
where  $y$ represents  the diameter of the volumes
modeled by spheres.
We now insert in the fundamental
Equation (\ref{fundamental}) for the chord  length
a PDF for the diameters as given by the previous equation.
The resulting integral,   which
models the  NPVT chords, is
\begin{equation}
g_{NPVT}(l)=
\frac
{
l{\pi }^{8/3}\sqrt [3]{3} \left( 3+\pi \,{l}^{3} \right)
}
{
5\,\Gamma  \left( 2/3 \right) {\pi }^{2}\sqrt [3]{{{\rm e}^{\pi \,{l}^
{3}}}}
}
\quad .
\label{GLBNONPOISSONIAN}
\end{equation}
We now make  
the following translation
\begin{equation}
U = L+ a \quad .
\end{equation}
The new translated PDF is  $g_{NPVT,1}(u;a)$ 
and the 
new constant of normalization is 
\begin {eqnarray}
g_{NPVT,2}(u;a)= C_{NPVT}g_{NPVT,1} (u;a) , \quad where \\
C_{NPVT}=\frac{1}{\int_0^{6-a}g_{NPVT,1}(u;a) du}
\end{eqnarray}
The  transformation of scale is
\begin{equation}
R = \frac{U}{b} \quad ,
\end{equation}
and the definitive  PDF for NPVT chords is
\begin{equation}
g_{NPVT,3}(r;a,b) =
\frac{g_{NPVT,2} (\frac{u}{b};a)}{b} \quad .
\label{GLBNONPOISSONIANSHIFT}
\end{equation}

\begin{figure*}
\begin{center}
\includegraphics[width=10cm]{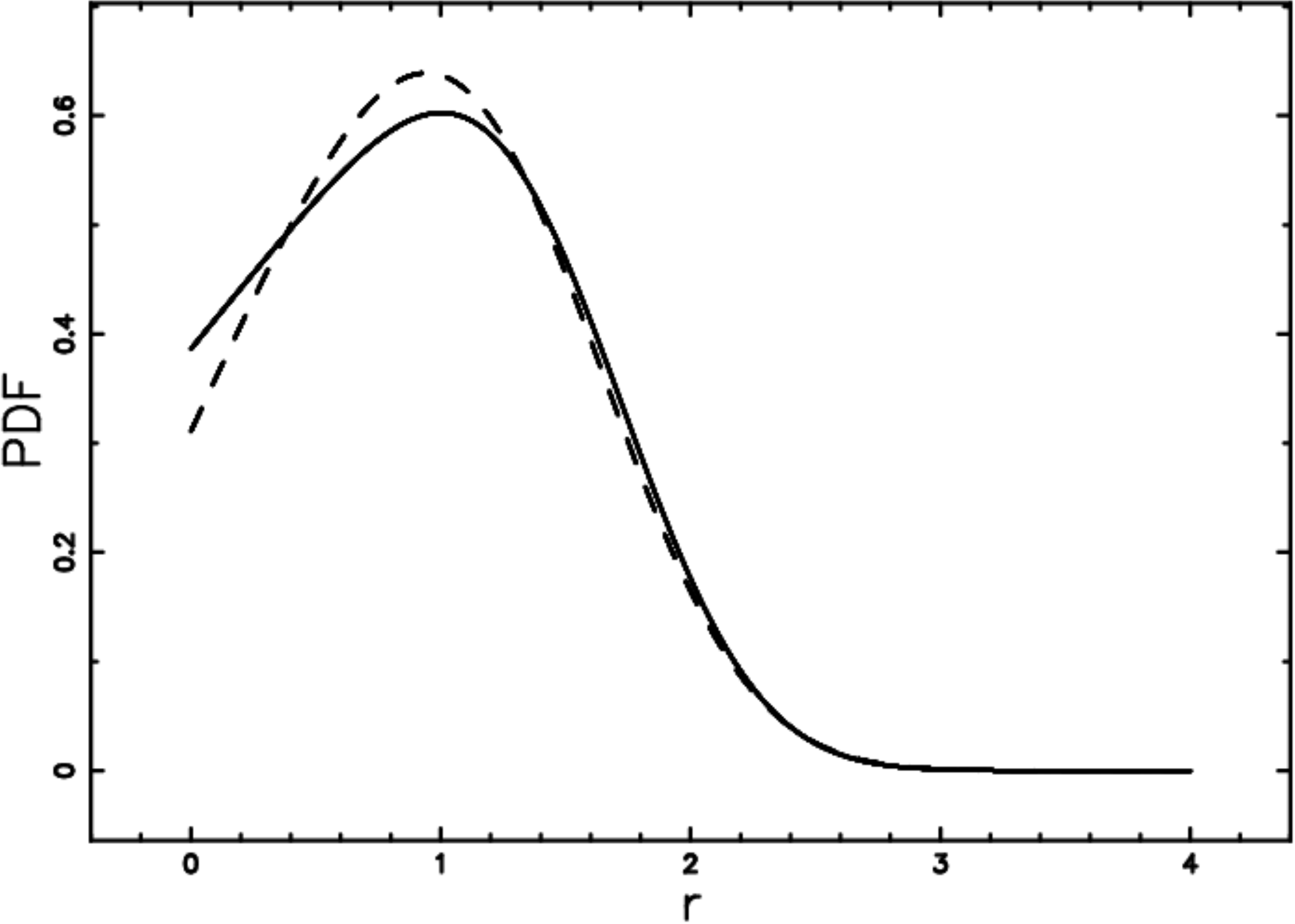}
\end {center}
\caption{
PVT PDF $g_3 (r;a,b) $ for chord length as a function of
$r$, when $b=2.452$, $a=0.563$
(full line),
and  NPVT PDF  $g_{NPVT,3} (r;a,b) $ 
when $b=1.741$, $a=0.36$
(dashed line).
In both cases  $<r>$=1.
}
\label{corda_pdf_c2}
    \end{figure*}

The behavior  of the NPVT $g_{NPVT,3} (r;a,b) $   is shown in
Figure  \ref{corda_pdf_c2}.
Table~\ref{table_parametersnpvt}
shows the average diameter, variance, 
skewness,  and kurtosis of the
already derived NPVT  $g_{NPVT,3}(r;a,b) $.
\begin{table}
 \caption[]
{
The parameters   of  \lowercase{$g_{NPVT,3} (r;a,b)$},
Eq.~(\ref{GLBNONPOISSONIAN}), relative to
the  NPVT  case when \lowercase{$b= 1.153, a=0.36$}. 
}
 \label{table_parametersnpvt}
 \[
 \begin{array}{ll}
 \hline
Parameter       & value   \\ \noalign{\smallskip}
 \hline
 \noalign{\smallskip}
Mean            & 0.662   \\
\noalign{\smallskip}
\hline
Variance        & 1.153  \\
\noalign{\smallskip}
\hline
Skewness        &  0.324  \\
 \hline
Kurtosis        &   -0.442  \\
 \hline
 \end{array}
 \]
 \end {table}
The resulting distribution function with scale will be
\begin{equation}
DF_{NPVT}(r:a,b) =\int_0^r g_{NPVT,3} (r;a,b) dr \quad .
\label{dfnonpoisson}
\end{equation}

\begin{figure*}
\begin{center}
\includegraphics[width=10cm]{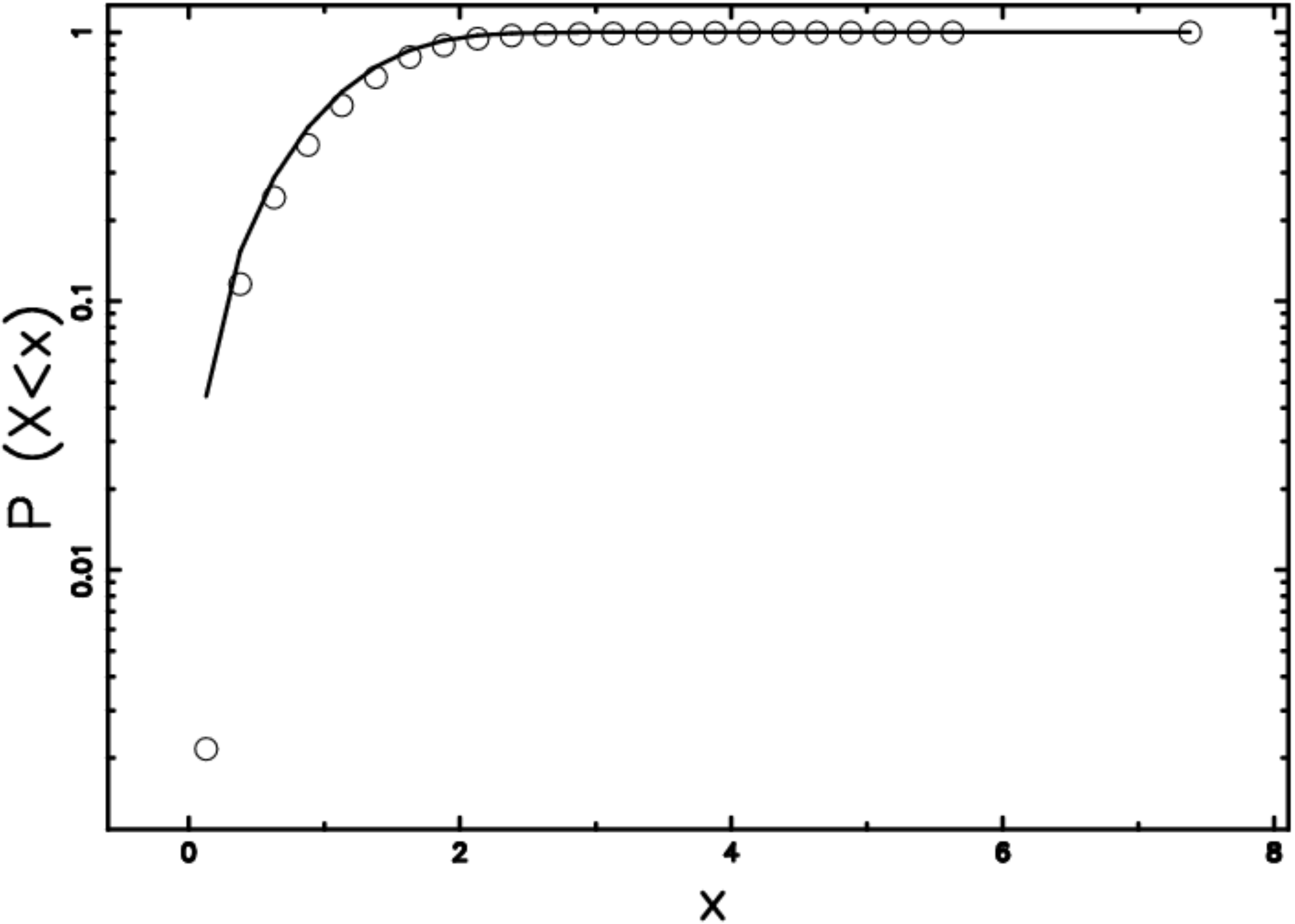}
\end {center}
\caption
{
Comparison between data
(empty circles) and theoretical curve
(continuous line)
of  the chord length distribution
when
$b=1.741$, $a=0.36 $, which 
means $<r>$ =1;  NPVT  case.
The maximum distance  between the two curves
is $d_{max}=0.03$.
The fraction of  volume forbidden to the NPVT seeds  
is $f= 16\%$.
}
\label{corda_df_npvt}
    \end{figure*}

Also in this case we produce  $900 000$ 3D
NPVT  cells   and we process
$9947$ chords  which
were obtained by adding together the results of
$40$ triples of  mutually perpendicular  lines.
The  numerical distribution of Voronoi lines  is
shown in Figure  \ref{corda_df_npvt}  with
the  display of  $DF_{NPVT}(r:a,b)$.

\section{Astrophysical applications}

\label{astrophysicalsec}
At the moment of writing, it is not  easy  to check
the already found PDFs for the chord distribution
in the local Universe.
This is because research  has been focused
on the intersection  of the maximal  sphere
of voids between  galaxies with the mid-plane
of a slab of galaxies, as an example
see  Figure 1 in \citet{Vogeley2012}.
Conversely, the  organization of the observed patterns
in slices of galaxies in irregular  pentagons
or hexagons having the property
of a tessellation has not yet  been developed.
We briefly  recall that in order to compute
the length  in a given direction of a slice of galaxies
the boundary between  a region and another region
should be clearly  computed in a digital  way.
In order to find  the value  of scaling, $b$, which models
the astrophysical  chords, some  first approximate methods will be suggested.
A {\it first} method  starts from the average chord 
for monodispersed bubble size distribution (BSD)
which are bubbles of constant radius $R$,
see (\ref{monogeometrical}), and  approximates it 
with 
\begin{equation}
<l>  = \frac{4}{3} <R>
\quad  ,
\label{quattroterzi}
\end{equation}
where $R$ has  been  replaced by  $<R>$.
We continue  by inserting  $<R>=18.23 h^{-1}$\ Mpc,
which is  the effective radius in  SDSS DR7, see Table 6  in 
in \citet{zaninetti2012e}, and 
in this case  $b=42.32 h^{-1}$\ Mpc.
A comparison should be made with the scaling 
of the   
probability of
obtaining a cross-section of radius $r$, which is  $\frac{31.33}{h}$\ Mpc,
see \citet{zaninetti2012e}.
The PDF of the chord conversely has average   value 
given by the previous equation (\ref{quattroterzi}) and therefore
$b$ is 4/3 bigger.

We now report a pattern of NPVT series of chords superposed
on a astronomical slice and
a simulation of an astronomical slice in the PVT approximation.

\subsection{The ESP}

\begin{figure*}
\begin{center}
\includegraphics[width=10cm]{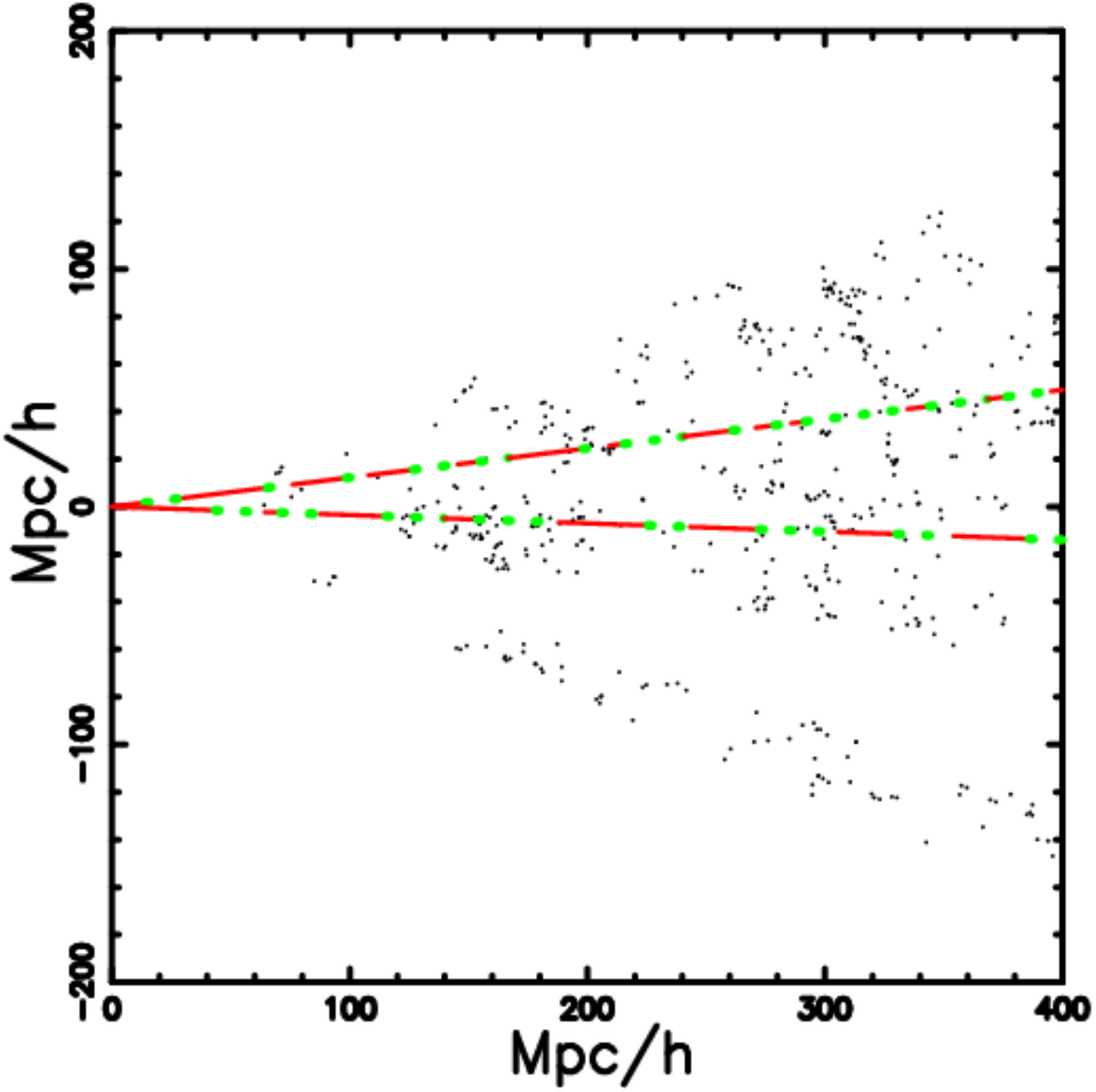}
\end {center}
\caption
{
Cone diagram of  ESP galaxies  when  the $X$ and $Y$ axes
are expressed  in Mpc and
 $H_{0}=100 $ $ \mathrm{\ km\ s}^{-1}\mathrm{\ Mpc}^{-1}$
(the Hubble constant),
which means $h$=1.
Two  lines in different directions  are drawn
with the lengths of the chords randomly generated
with  NPVT PDF  $g_{NPVT,3} (r;a,b)$
when $b=42.32 h^{-1}$\ Mpc and $a=0.36$.
Each chord has a different color,
red and green,
and different line styles,
full and  dotted.
}
\label{esoslice}
    \end{figure*}

As an example, we fix attention on the Eso Slice
Project (ESP)
which covers a strip of 22(RA) $\times$ 1(DE) square degrees,
see \citet{Vettolani1998}.
On the ESP we report two lines over which a random distribution
of chords follows the NPVT equation (\ref{GLBNONPOISSONIAN})
with $b=42.32 h^{-1}$\ Mpc
and $a=0.36$,
see Figure \ref{esoslice}.

\subsection{A simulated slice}

\begin{figure*}
\begin{center}
\includegraphics[width=10cm]{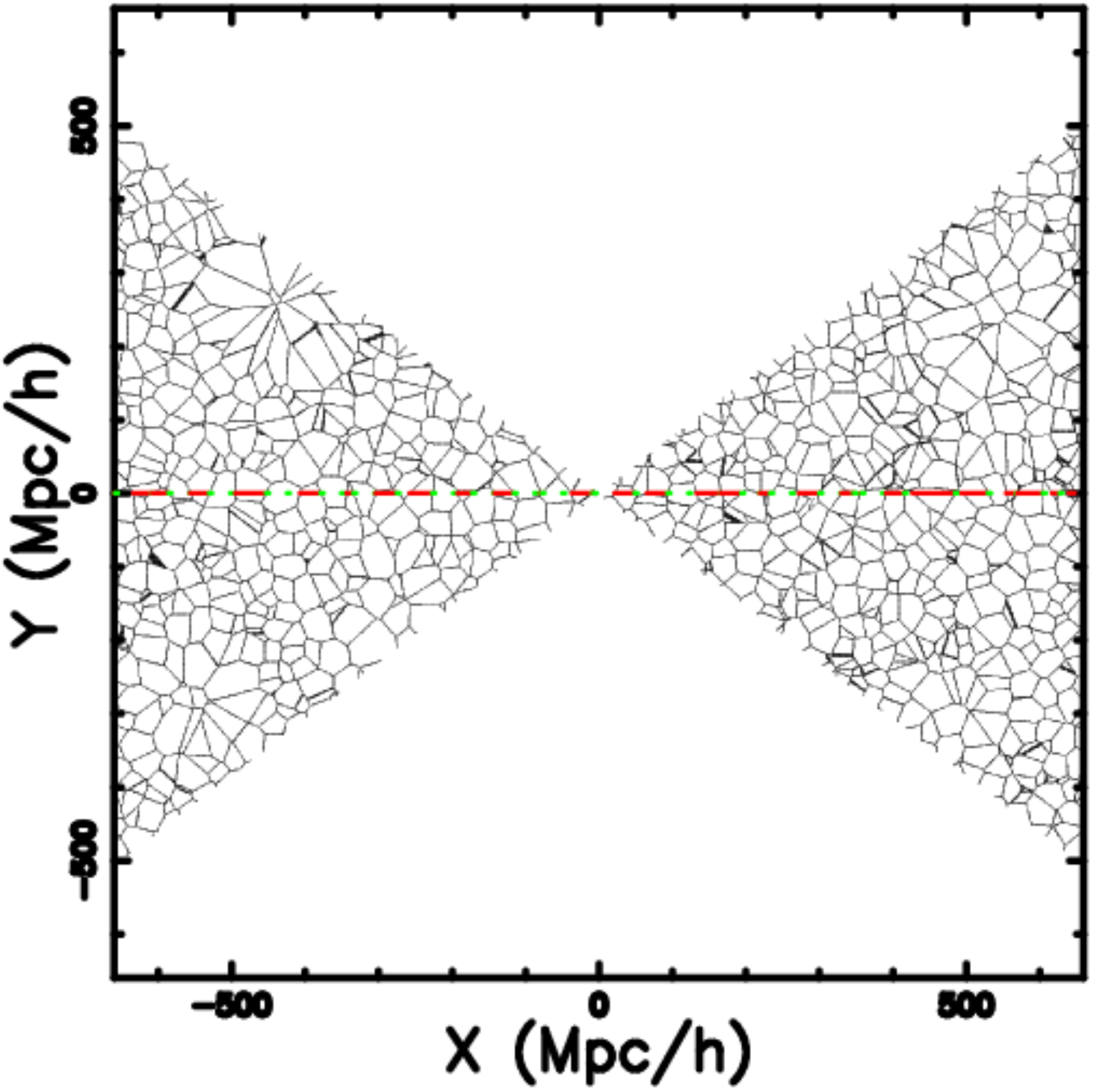}
\end {center}
\caption
{
Portion of the  NPVT  $V_p(2,3)$;
cut on the  X-Y plane   when two strips of
$75^{\circ}$ are considered.
The  parameters  of  the simulation,
see \citet{zaninetti2010a},
are       pixels   = 1500,
  side     = 131 908 Km/sec
  and     amplify  = 1.2.
Along a  line which crosses the center,  the chords between a face
and the next one  are drawn.
Each chord has a different color, red and green,
and different line styles, full and  dotted.
The fraction of  volume forbidden to the NPVT seeds  
is $f= 16\%$  and the  number of the seeds, 77 730, is chosen in order to 
have $b=42.32 h^{-1}$\ Mpc.
}
          \label{cut_middle_color}%
    \end{figure*}

In order to simulate  the 2dF Galaxy Redshift Survey (2dFGRS),
we simulate     a 2D cut on a 3D PVT network  organized
in   two  strips
about $75^{\circ}$ long,
see  Figure \ref{cut_middle_color}

\begin{figure}
\begin{center}
\includegraphics[width=10cm]{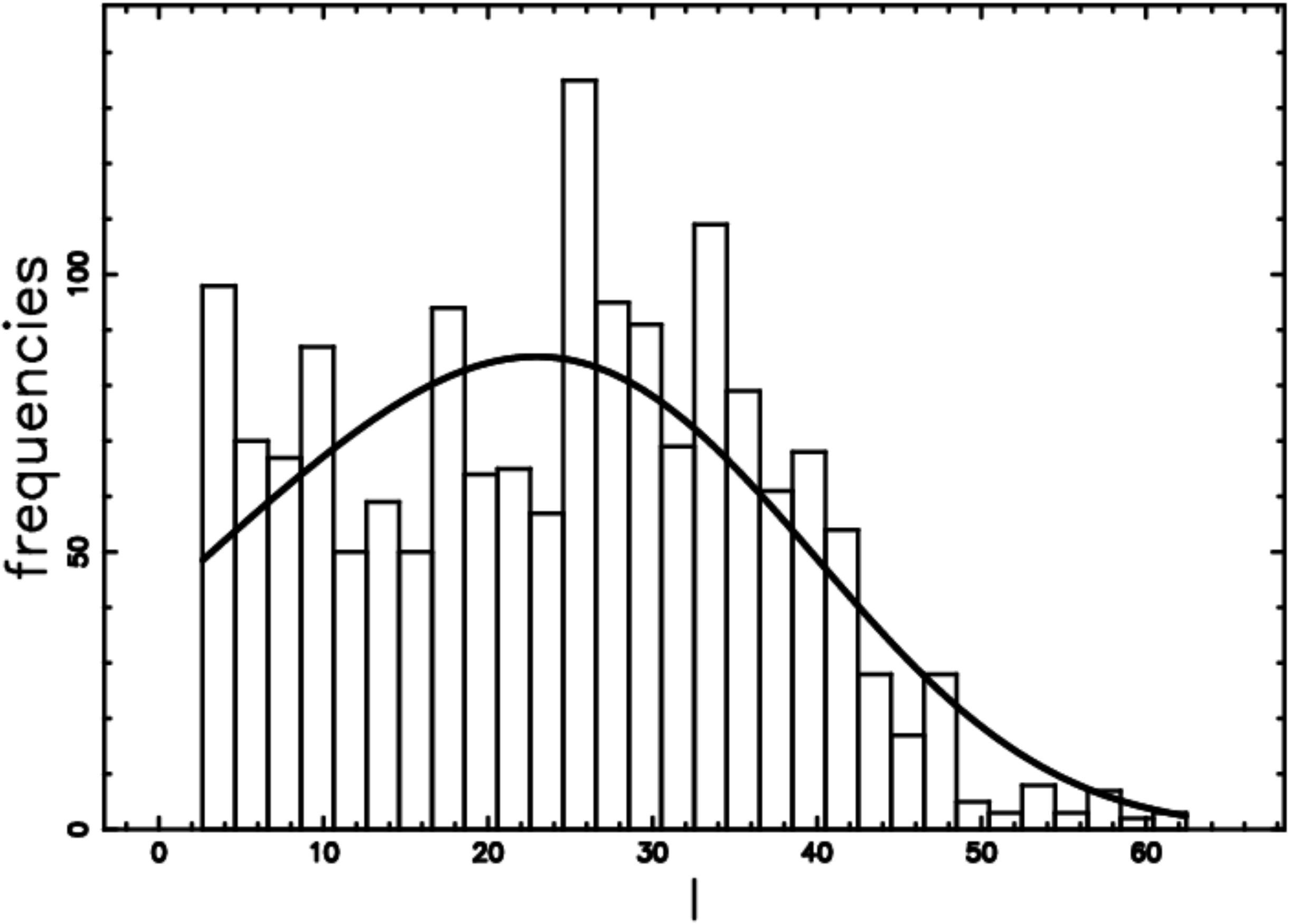}
\end {center}
\caption
{
Histogram of
the
frequencies  in chord lengths
along 31 lines (1617 sequential chords) with a superposition of the
theoretical NPVT PDF
as represented by Eq.~(\ref{GLBNONPOISSONIAN}).
The number of bins is 30, the reduced $\chi^2$ is 9.21, and 
$b=42.32 h^{-1}$\ Mpc.
}
\label{corda_taglio}
    \end{figure}

\begin{figure}
\begin{center}
\includegraphics[width=10cm]{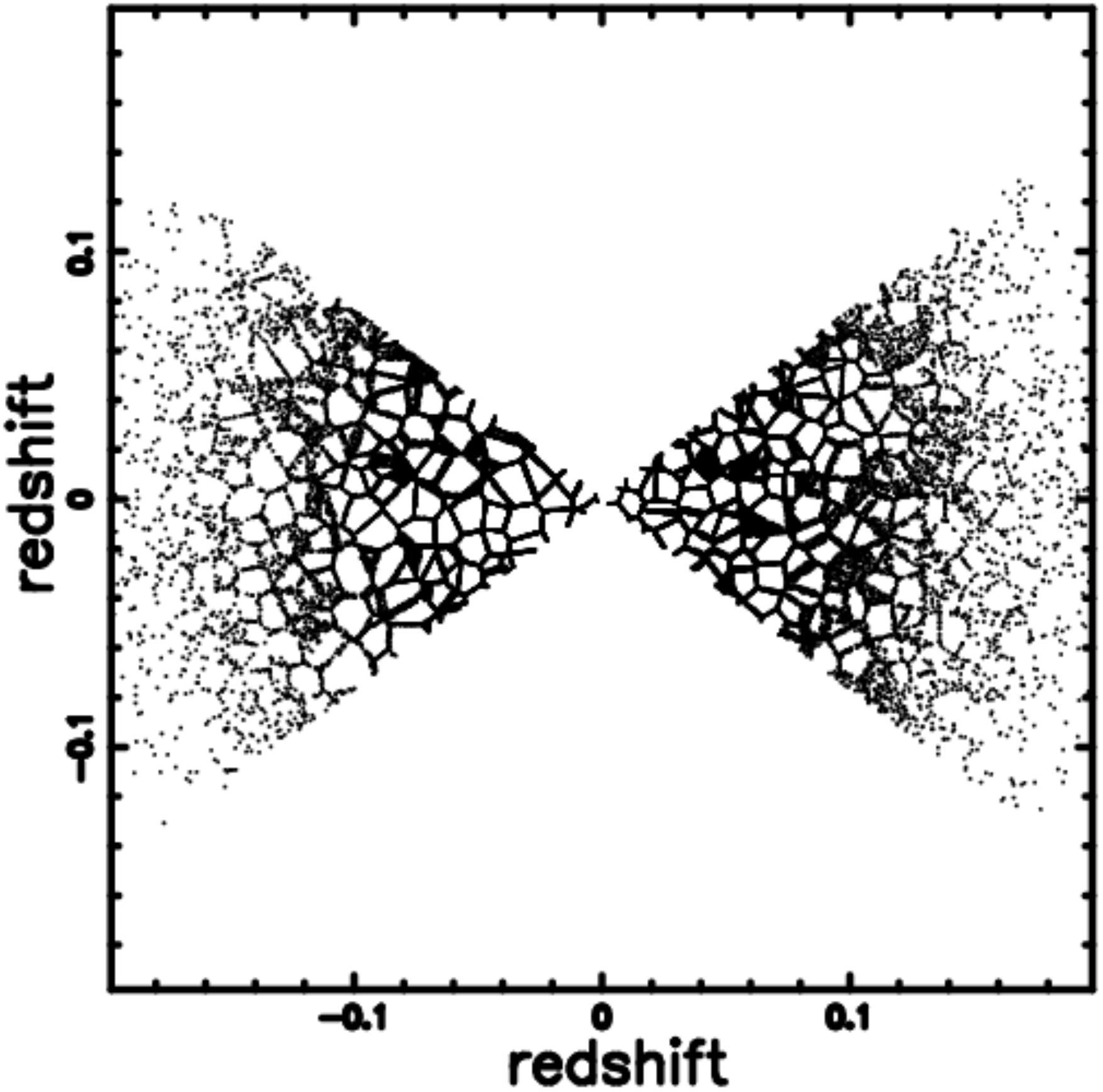}
\end {center}
\caption
{
Polar plot
of the  pixels  belonging to a
slice   $75^{\circ}$~long  and $3^{\circ}$
wide generated by NPVT seeds.
This plot contains  40 000 
galaxies,
the maximum frequency of theoretical
galaxies is at  $z=0.056$.
In this plot, $\mathcal{M_{\sun}}$ = 5.33   and $h$=1.
}
\label{voro_2df_cones}
    \end{figure}

Figure \ref{corda_taglio} shows  a superposition
of the numerical  frequencies
in chord lengths
of a 3D PVT simulation  with 
the curve of the theoretical PDF of PVT chords,
$ g(l,b)$,  as  given by Eq.~(\ref{GLBPOISSONIAN}).
The  distribution of the galaxies as  
given by NPVT 
is reported in Figure~\ref{voro_2df_cones},
conversely  Figure 15  in
\citet{zaninetti2010a} was  produced with PVT. 

\section{Conclusions}

A  mathematical  method  was developed 
for  the chord length  distribution obtained
from the sphere diameter distribution function
which approximates the volume of the PVT.
The  new chord length  distribution as  represented
by  the PVT formula  (\ref{GLBPOISSONIAN})
can give  mathematical support
for  the periodicity along  a line, or a cone,
characterized  by a small solid angle.
At the same time, a previous  analysis has  shown
that  the best fit  to the effective  radius  of the cosmic
voids as  deduced from
the  catalog SDSS R7  is  represented  by a Kiang function
with $c \approx 2$.
Also in this case
we found an expression  for the chord length
given  by the 3D  NPVT  formula
(\ref{GLBNONPOISSONIAN}).
In order to produce this kind of NPVT volumes,   a new type of seed 
has  been introduced, see  Section  \ref{secnewseeds}.
These new seeds are also used to simulate the intersection 
between a plane and the NPVT network.
A careful choice of the number of seeds allows matching the simulated
value of the scaling $b$ with the desired value of 
$b=42.32 h^{-1}$\ Mpc, see Figure \ref{corda_taglio}.

\hrulefill

\end{document}